\documentclass[aps,showpacs,twocolumn]{revtex4}
\usepackage{graphicx}

\begin{document}

\title{Superconducting proximity effect in the presence of strong spin scattering}
\author{J. Eom,$^1$ J.\ Aumentado,$^2$ V. Chandrasekhar,$^2$ P.M. Baldo,$^3$ and L.E. Rehn$^3$}
\affiliation{$^1$Department of Physics, Sejong University, Seoul, 143-747 Korea}
\affiliation{$^2$Department of Physics and Astronomy, Northwestern University, Evanston, IL
60208}
\affiliation{$^3$Materials Sciences Division, Argonne National Laboratory, Argonne, IL}

\date{\today}

\begin{abstract}
We report measurements of the four terminal temperature dependent
resistance of narrow Au wires implanted with ~100 ppm Fe impurities in
proximity to superconducting Al films.  The wires show an initial
decrease in resistance as the temperature is lowered through the
superconducting transition of the Al films, but then show an
increase in resistance as the temperature is lowered further.  In contrast to the case of pure Au
wires in contact with a superconducting film, the resistance at
the lowest temperatures rises above the normal state resistance.
Analysis of the data shows that, in addition to contributions from magnetic scattering and electron-electron interactions, the temperature dependent resistivity shows a substantial contribution from the superconducting proximity effect, which exists even in the presence of strong spin scattering.
\end{abstract}

\maketitle

The properties of a normal metal in contact with a superconductor
are modified by the superconducting correlations induced by the
proximity effect.  Microscopically, these correlations arise from
phase coherent Andreev reflection of electrons in the normal metal
at the normal metal/superconductor (NS)
interface~\cite{Andreev64}.  In diffusive normal-metal samples whose
dimensions are less than or comparable to the electron phase
coherence length $L_{\phi}$,  these correlations give rise to a non-monotonic
dependence of the resistance as a function of temperature below
the transition temperature $T_c$ of the superconductor, the so-called reentrance effect.
The resistance  initially decreases as the
temperature is lowered, reaches a minimum at some intermediate
temperature $T_{min}$, and then \emph{increases} as the
temperature is lowered still further.  

The temperature $T_{min}$
at which the minimum is observed is determined by the Thouless
energy $E_c =\hbar D / L^2$, where $D = v_F
\ell/3$ is the electron diffusion constant, $v_F$ the Fermi
velocity, $\ell$ the elastic mean free path in the normal
metal, and $L$ is the length of the normal wire \cite{Golubov97}.  In the absence of electron-electron interactions, the resistance is expected to approach its normal state value $R_n$ at $T=0$.  In the presence of electron-electron interactions, the resistance can be either less or greater than $R_n$, depending on whether the attractions are attractive or repulsive \cite{Nazarov96,Zhou95}.  Experimentally, reentrant behavior has been observed in many experiments on NS and
superconductor/semiconductor (N-Sm)
structures~\cite{Charlat96,Hartog96}. 
Theoretically, the presence of long-range electron phase coherence is thought to be 
critical for the existence of the superconducting proximity effect.  
The electron phase coherence in the normal metal
$L_{\phi}$ provides an upper cutoff for penetration of
superconducting correlations into the normal metal \cite{Nazarov96}.  Consequently, in systems in
which $L_{\phi}$ is short, the proximity effect is expected to be
correspondingly suppressed.  

In this paper, we report on measurements of the temperature
dependent resistance $R(T)$ of Au wires with ~100 ppm magnetic Fe
impurities in contact with an Al film.  The wires are designed in a
four-terminal configuration so only the normal metal wire itself (and not the NS
interface) is measured.  The wires show an initial
decrease in resistance as the temperature is lowered through the
superconducting transition of the Al films. As the temperature is
lowered still further, the resistance of the AuFe wires starts to
increase, and finally rises above the normal state
resistance at the lowest temperatures.  Analysis of the Kondo and
electron-electron interaction contributions to the resistivity show
that a substantial contribution to $R(T)$ is associated with the
superconducting proximity effect, in spite of the fact that the presence of the magnetic impurities is expected to drastically attenuate $L_{\phi}$.   

\begin{figure}
%\vspace{-4cm}
\includegraphics[width=7.2cm]{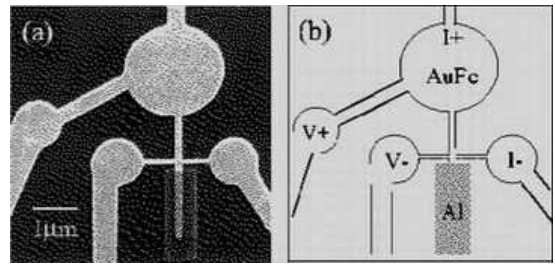}
%\vspace{-4.5cm}
\caption{(a) Micrograph of a typical AuFe wire in proximity to a
Al film. (b) Schematic of probe configuration.}
\label{fig1}
\end{figure}

% Sample fabrication and details of the sample
The two AuFe wires discussed in this paper were patterned by multilevel
electron beam lithography onto oxidized Si substrates.  They form part
of a sample designed to investigate the size dependence of the
thermopower in the Kondo
effect.  These results will be discussed in a later publication; here
we concentrate on the temperature dependent resistance of the two AuFe
wires that were used as thermometers in the experiment. In the
first level of lithography, a 51-nm-thick film was deposited by thermal
evaporation from a 99.999\%-pure Au target. After this, the entire Au film was implanted by Fe ions to a
concentration of $\sim$ 100 ppm. Finally, a 75-nm-thick Al film
was deposited as the superconductor.
Figure 1(a) shows scanning electron micrograph of one of the
AuFe wires.  The two wires had lengths of 0.5
$\mu$m and 1.0 $\mu$m, and a width of 160 nm.  Four leads attached to
the wires enabled us to perform four-terminal resistance measurements
on the normal metal by itself (see the schematic if Fig. 1(b)); a
fifth lead was overlaid with a 75 nm thick Al film to induce a
proximity effect in the wire.  The resistivity of the Au film at $T=$1 K was  $\rho_0=$ 3.52 $\mu \Omega$-cm, and the thermal length $L_{T} =
\sqrt{\hbar D / k_{B}T} \approx$ 0.29 $\mu$m at 1 K. The
measurements were performed in a dilution refrigerator and all
electrical terminals were carefully shielded to
minimize unwanted noise effects.

% Results and discussion

Figures 2(a) and 2(b) show $R(T)$ of AuFe wires of length 0.5 $\mu$m and 1.0 $\mu$m respectively.  Both wires show a sharp drop in resistance at approximately 1.2 K, the temperature $T_c$ at which the Al films in contact with the wires go through their superconducting transitions.  It should be noted again that, due to the four-terminal nature of the measurement, this drop in resistance at $T_c$ is associated with the Au wire, and not an inadvertent measurement of the resistance of the Al film.   This indicates that the drop in resistance below $T_c$ that we observe is indeed associated with the superconducting proximity effect.  

Below $T\simeq$0.8 K, the resistance of both wires increases rapidly, rising above the normal state resistance at lower temperatures.  At the lowest temperatures, the resistance eventually decreases, resulting in a small peak at $T\simeq$60 mK.  To discuss the possible origins of this temperature dependence, it is instructive to plot the data for the two wires in terms of changes in the resistivity from the normal state resistivity.  These data are shown in Fig. 3.  Plotted in this manner, it can be seen that the resistivity changes for the two wires differ substantially only below 0.8 K.

\begin{figure}
\includegraphics[height=11.5cm]{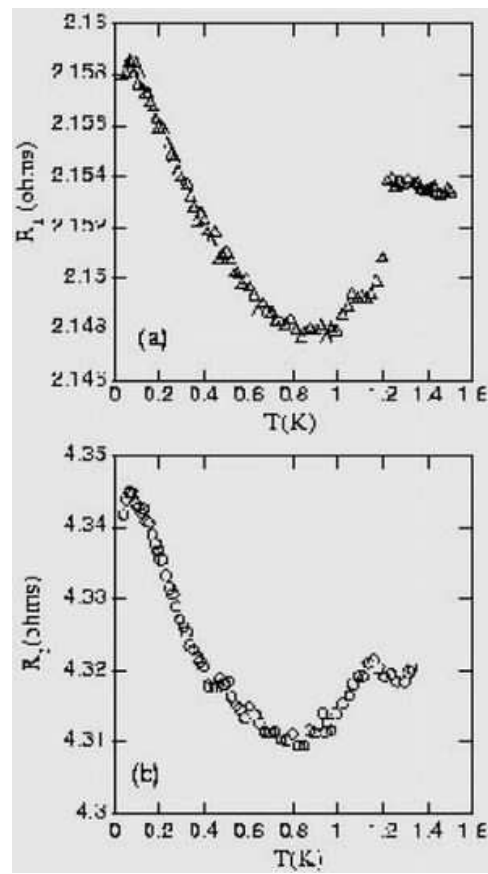}
\caption{(a) Resistance of the 0.5 $\mu$m long AuFe wire, as a function of temperature (b) Resistance of the 1.0 $\mu$m long wire, as a function of temperature.}
\label{fig2}
\end{figure}

In addition to the proximity effect, three other effects may contribute substantially to the low temperature $R(T)$: weak localization (WL), electron-electron interactions (EEI), and scattering by magnetic impurities.  In one dimension. the functional form of the WL contribution to the resistivity can be written as \cite{altshuler} 
\begin{equation}
\Delta \rho_{WL} = \frac{\rho_0^2}{\pi \hbar/e^2} \frac{L_{\phi}}{A}
\label{WL}
\end{equation}
where $A=8.16 \times 10^{-11}$ cm$^2$ is the cross-sectional area of the wires.  (Since we are interested here only in estimating in the total size of the contribution, we ignore the effects of spin-orbit scattering.)  The temperature dependence of the WL contribution is determined by the temperature dependence of $L_{\phi}$, which in turn is determined by the temperature dependence of the magnetic scattering time $\tau_s$.  Although there has been a tremendous amount of interest recently in the role of magnetic scattering in electron coherence \cite{lin}, for our discussion, we are interested only in an upper bound on the WL localization contribution.  We therefore choose $L_{\phi}\simeq1$  $\mu$m as an upper limit over the temperature range of interest, a value is consistent with recent experiments on WL in AuFe wires with concentrations comparable to those in our samples \cite{schopfer}.  Using values for our samples, this gives an upper limit of $\Delta \rho \simeq$ 1 n$\Omega$-cm for the total change in resistivity due to WL below $T=$ 0.8 K.  This is a small contribution to the total change in resistivity shown in Fig. 3.

\begin{figure}
\includegraphics[height=10cm]{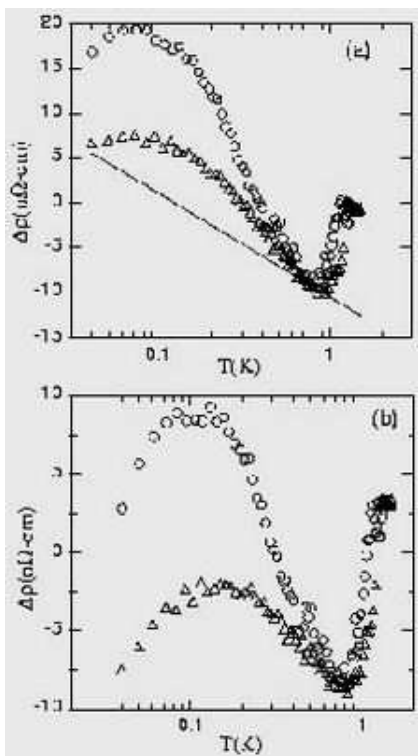}
\caption{(a)  Data of the two wires of Fig. 2, plotted in terms of change of resistivity.  Open triangles, 0.5 $\mu$m long wire; open circles, 1.0 $\mu$m long wire.  Solid line is a fit to the Kondo resistivity, Eq. (3), with a Kondo slope of 0.12 n$\Omega$-cm per ppm  per decade of temperature, as discussed in the text.  (b)  Data for the two wires as in (a), with the Kondo resistivity represented by the solid line in (a) subtracted from both curves.}
\label{fig3}
\end{figure}

The one-dimensional form of the EEI contribution can be written in a form similar to Eq.(1): \cite{Altshuler85,chandrasekhar}
\begin{equation}
\Delta \rho_{ee} = \alpha \frac{\rho_0^2}{\pi \hbar/e^2} \frac{L_T}{A}
\label{EEI}
\end{equation}
where $\alpha$ is a parameter of order unity, which has been found to be approximately 1.5 for Au \cite{chandra2}.  Again, the total change in resistivity due to electron-electron interactions is only  $\Delta \rho \simeq$ 1 n$\Omega$-cm.  Consequently, the contributions to the resistivity due to either WL or EEI effects is negligible in our experiment, and will be ignored in our analysis.

The last contribution of importance is associated with scattering by magnetic impurities.  For dilute concentrations of magnetic impurities, where the magnetic interactions between impurities can be ignored, scattering of electrons off the magnetic impurities gives rise to a contribution to the temperature dependent resistance, the so-called Kondo effect.  At temperatures far above the characteristic Kondo temperature $T_K$ of the specific metal/magnetic-impurity system, the Kondo effect gives rise to a logarithmic contribution to the temperature dependent resistivity\cite{Kondo}:
\begin{equation}
\Delta \rho_K \simeq - \rho_K \ln(T/T_K)
\label{Kondo}
\end{equation}
where $\rho_K$ is a constant for a specific metal/magnetic-impurity system, and has been determined to be 0.11-0.17 n$\Omega$-cm per decade of temperature change per ppm of Fe impurities in a Au host \cite{Loram70}.  At lower temperatures, in the unitarity limit of the Kondo effect, the resistance due to magnetic scattering is expected to approach a constant value quadratically with the temperature $T$ as $T \rightarrow 0$ \cite{nozieres}.  However, this single impurity limit behavior is frequently masked by the effect of RKKY interactions between the magnetic impurities, which tend to freeze the magnetic impurity spins in random orientations as the temperature is reduced below a characteristic temperature $T_{f}$ \cite{daybell}.  Below $T_{f}$, the resistance of the metal decreases as the magnetic scattering is frozen out. 

The $\Delta \rho(T)$ curves for both wires in Fig. 3 show behavior that is at first sight consistent with magnetic scattering, i.e., the resistance first increases as the temperature is decreased below $T\simeq0.8$ K, reaches a maximum at around $T\simeq60$ mK, and then decreases below this temperature.  However, if this temperature dependence is solely due to scattering by magnetic impurities, then, according to Eq. (3), the $\Delta \rho(T)$ curves for both wires should be the same, since both wires have the same magnetic impurity concentration; indeed, they are part of the same sample, separated by a distance of less than 5 microns.  In fact, the slope of $\Delta \rho(T)$ for the 1 $\mu$m long wire between 0.2 K and 0.8 K is about a factor of 1.65 larger than the corresponding slope for the 0.5 $\mu$m long wire, and the slope for even the 0.5 $\mu$m long wire is a factor of two larger than expected for scattering by only magnetic impurities.

For a quantitative estimate of the contribution to the resistivity due to the Kondo effect, we choose a value of $\rho_K$ in Eq. (3) to match the slope of the resistivity of both Au wires above $T_c$, where the contribution from the superconducting proximity effect is absent, obtaining a value of $\Delta \rho_K$ = 0.12  n$\Omega$-cm per decade of temperature change per ppm.   This value is in good agreement with previous work on ion-implanted AuFe wires of similar concentration \cite{chandra2}.  The solid line in Fig. 3 shows the resulting logarithmic contribution due to the Kondo effect using this value of the slope.  The contribution due to the Kondo slope is clearly much less than the resistivity change in either wire.  Of course, this simple analysis does not capture the reduction in resistivity at very low temperatures, which arises from suppression of magnetic scattering due to the incipient formation of a spin-glass.  Theoretical models have been developed that take into account both the Kondo and spin-glass effects \cite{larsen}; these models predict that there is a universal function that describes the resistivity for samples of a particular metal/magnetic-impurity system with a specific impurity concentration.  However, there is some question as to the applicability of these models to disordered mesoscopic samples \cite{neuttiens}, and in any case, the presence of the superconducting proximity effect makes a quantitative comparison to this theory difficult.  We shall therefore focus on a qualitative discussion of the low temperature resistivity of the wires, attempting to draw some general conclusions about the superconducting proximity effect in the presence of magnetic impurities.

In terms of resistivity,  the Kondo and spin-glass contributions due to magnetic scattering should be the same for both wires.  Consequently, the difference between the resistivity of the two wires shown in Fig. 3(a) arises primarily from the superconducting proximity effect (we have noted earlier that the WL and EEI contributions are small).  In order to emphasize this difference, we show in Fig. 3(b) the data for the two wires of Fig. 3(a), with the logarithmic Kondo contribution represented by the solid line in Fig. 3(a) subtracted.  Unfortunately, unlike the WL or EEI contributions, there is no simple analytical expression to determine the contribution of the superconducting proximity effect.  Estimating this contribution would require a numerical solution of the Usadel equation for our particular geometry.  However, one can qualitatively compare the temperature dependence seen in Fig. 3(b) to the behavior expected for a normal metal wire in contact with a superconductor. 

For a wire of length $L<L_\phi$ connected to a superconducting reservoir by a perfectly transparent interface on one end, and a normal-metal reservoir on the other,  the overall change in resistance is typically $\simeq 0.1 R_n$.  This value is reduced as the transparency of the NS interface decreases, although the temperature $T_{min}$ at which the minimum in resistance is observed does not change appreciably. 
$L_{\phi}$ acts as an upper length scale for the proximity effect; if $L>L_{\phi}$, the effective length of the wire can be considered to be $L_{\phi}$.  For a constant $L_\phi<L$, this has two consequences in comparison to the case $L_\phi>L$.  First,  $T_{min}$ increases, and second, the overall change in resistance decreases, since now only a fraction $L_\phi/L$ of the wire contributes to the proximity effect.  $L_\phi$ in our samples is expected to be shorter in comparison to normal metals with no magnetic impurities, due to the enhanced dephasing introduced by them.  The overall change of $0.5 \%$ of $R_n$ at low temperatures observed in our samples is consistent with resistance changes observed in previous experiments on proximity-coupled normal metals, if $L_\phi$ is short, but not negligible.

If $L_\phi$ is less than $L$ and itself varies with temperature,  the temperature dependence of the proximity effect might be expected to be far more complicated.  To our knowledge, this regime has not been discussed in detail theoretically  or experimentally \cite{black}.  Experimentally, of course, $L_\phi$ is in general temperature-dependent.  In the absence of magnetic impurities, $L_\phi$ is observed to increase as a power law with decreasing $T$, eventually saturating at low temperatures in some experiments \cite{lin}.  In the presence of strong magnetic scattering, $L_\phi$ may have a non-monotonic temperature dependence, associated with the peak in spin-flip scattering at $T_K$ due to the Kondo effect \cite{peters,cvh,mohanty,schopfer}.  Thus, the temperature dependence of the proximity effect contribution to the resistance may directly reflect the temperature dependence of the mechanisms that contribute to electron decoherence in disordered normal metals, both with and without magnetic impurities.  Consequently, although it is tempting to ascribe the increase in resistance above $R_n$ that we observe in our samples to the presence of electron-electron interactions as described in the conventional theory of the proximity effect, we believe that a full understanding of the low temperature resistance anomalies requires further theoretical and experimental investigation.

In summary, we have measured the temperature dependent resistance of 100 ppm AuFe wires that are in close contact with a superconducting film.  In addition to the contribution due to scattering from magnetic impurities, a significant contribution to the low temperature resistance comes from the superconducting proximity effect, which exists even in the presence of strong spin scattering.  The temperature dependence of this latter contribution is unusual, and we believe it arises from the temperature dependence of the underlying mechanisms that contribute to the loss of electron phase coherence in disordered normal metals.  

This work was supported by the NSF through DMR-0201530, and by the Korea Research Foundation Grant
(KRF-2001-003-D00051), and the US Department of Energy under Contract No. W-31-109-Eng-38.

%-----------References---------------------------------------

%------------FIGURE CAPTIONS---------------------------------


\begin{thebibliography}{text}
\bibitem{chandrasekhar}  The equation in  this form is given in \cite{chandra2}.
\end{thebibliography}
\end{document}